\documentstyle[12pt]{article}
\begin{document}
\title{ \Large\bf The Physical Spectrum of Conformal SU(N) Gauge
 Theories } \author{Thomas Appelquist\thanks{Electronic address:
 twa@genesis3.physics.yale.edu}\ \ and Francesco
 Sannino\thanks{Electronic address:
 sannino@apocalypse.physics.yale.edu}\\ \\ {\it Department of Physics,
 Yale University, New Haven, CT 06520-8120}} \date{\today} \maketitle
\begin{picture}(0,0)(0,0)
\put(346,226){YCTP-P12-98}
\put(346,210){hep-ph/9806409}
\end{picture}
\vfill
\begin{abstract}
We investigate the physical spectrum of vector-like
 $SU(N)$  gauge theories with infrared coupling close to but above 
the critical value  for a conformal phase transition.  We use
dispersion relations, the momentum dependence of the dynamical
fermion mass and resonance saturation.  We show that the second
spectral function sum rule is substantially affected by the continuum
contribution, allowing for a reduction of the axial vector - 
vector mass splitting with respect to QCD-like
theories.  In technicolor theories, this feature can result in 
a small or even negative contribution to 
the electroweak $S$ parameter.
\end{abstract}
\vfill
\begin{flushleft}
\footnotesize
PACS numbers:11.15.-q, 12.60.Nz, 12.30.Rd
\end{flushleft}
\vfill
\newpage
\section*{}
 The past few years have seen renewed interest in the study of the
physical properties of gauge field theories with an infrared fixed point
\cite{IS}. While ordinary QCD doesn't fall into this category, other
theories of great interest do. The study of  $N=1$ supersymmetric 
gauge theories has lead to a reasonable
picture for the different phases depending on 
the number $N_f$ of matter multiplets
\cite{IS}. For a range of $N_f$, an infrared fixed point exists and the
theory is in the ``non-abelian coulomb phase''.  Even ordinary gauge theories
can contain infrared fixed points depending on the number of matter
fields. In this letter we consider an $SU(N)$ gauge theory with $N_f$
flavors, whose quantum symmetry group is
$\displaystyle{SU_L(N_f)\otimes SU_R(N_f) \otimes U_V(1)}$.  It is
well known that if $N_f$ is large enough, but below $11N/2$, an
infrared fixed point $\alpha_*$ exists, determined by the first two
terms in the renormalization group $\beta$ function.  For $N_f$ near
$11N/2$, $\alpha_*$ is small and the global quantum  symmetry group
remains unbroken.  For small $N_f$, on the other hand,  we expect the
chiral symmetry group $\displaystyle{SU_L(N_f)\otimes SU_R(N_f)}$ to
break to its diagonal subgroup.

 It is an important and unsolved problem to determine where the phase
transition takes   place as $N_f$ is varied. One possibility is that
it happens at a relatively large value of $N_f/N (\approx 4)$
corresponding to an infrared fixed point accessible in perturbation
theory \cite{ATW}.  An alternative possibility is that the transition
takes place in the strong coupling regime, corresponding to relatively
small values of $N_f/N$ \cite{mawhinney}.

The larger value emerges from studies
of the renormalization group improved gap equation, and corresponds to
the perturbative infrared fixed point  $\alpha_*$ 
reaching a certain critical value
$\alpha_c$. These studies also show that the order parameter, for
example the Nambu-Goldstone boson decay constant $F_{\pi}$, vanishes
continuously at the transition. A recent analysis \cite{AS} indicates 
that instanton
effects could also trigger chiral symmetry breaking at comparably large 
values of $N_f/N$.

Here, we will study the spectrum of states
in the broken phase near a large $N_f/N$ transition, in particular  the
possibility that  parity doubling or possibly even parity inversion takes
place. This could have important consequences for electroweak
symmetry breaking, since near-critical gauge theories provide a
natural framework for walking technicolor theories
\cite{postmodern}. The $S$ parameter, in particular, is sensitive to the
relative positioning of parity partners.

 It has been noted \cite{ATW} that in such a transition, there
are no light degrees of freedom in the symmetric phase other than the
fermions and gluons. In the broken phase near the transition, the
approximate conformal symmetry suggests that all massive states 
scale to zero with the order parameter
\cite{Sekhar}. It follows that a simple Ginzburg-Landau Lagrangian
cannot be used to explore the transition.

In a near-critical theory governed by an infrared
fixed point the coupling is near the fixed point for momenta
ranging from the small scale associated with the 
physical states up to some intrinsic, renormalization scale
$\Lambda$. In Reference~\cite{ATW}, this scale was defined such that
$\alpha_{\Lambda}\equiv \alpha(\Lambda) \approx 0.78 \alpha_*$. Above
this scale, asymptotic freedom sets in.

 A natural framework for exploring the relation between the fixed
point behavior and the spectrum of light states is provided by 
the Weinberg sum rules. The
relevant two point Green function is the time honored vector-vector
minus axial-axial vacuum polarization, known to be sensitive to chiral
symmetry breaking.  We define
\begin{equation}
i\Pi_{\mu \nu}^{a,b}(q)\equiv \int\!d^4x\, e^{-i qx}
\left[<J_{\mu,V}^a(x)J_{\nu,V}^b(0)> -  
 <J_{\mu,A}^a(x)J_{\nu,A}^b(0)>\right] \ ,
\label{VA}
\end{equation}
where
\begin{equation}
\Pi_{\mu \nu}^{a,b}(q)=\left(q_{\mu}q_{\nu} - g_{\mu\nu}q^2 \right) \,
\delta^{a b} \Pi(q^2) \ .
\end{equation}
Here $a,b$, $=1,...,N_f^2-1$, label the flavor
currents and the $SU(N_f)$ generators are normalized according to 
$\rm{Tr} \left[T^a T^b\right]= (1/2) \delta^{ab} $.  The
function $\Pi(q^2)$ obeys the unsubtracted dispersion relation
\begin{equation}
\frac{1}{\pi} \int_0^{\infty}\!ds\, \frac{{\rm Im}\Pi(s)}{s + Q^2}
=\Pi(Q^2) \ ,
\label{integral}
\end{equation}
where $Q^2=-q^2 >0$, as well as the constraint \cite{W}
$\displaystyle{-Q^2 \Pi(Q^2)>0}$ for $0 < Q^2 < \infty$.

Because the theory exhibits asymptotic freedom 
above $\Lambda$, the behavior of $\Pi(Q^2)$ at asymptotically high
momenta is the same as in   ordinary QCD, i.e. it scales like $Q^{-6}$
\cite{BDLW}. Expanding the left hand side of the dispersion   relation
thus leads to the two conventional spectral function sum rules
\begin{equation}
\frac{1}{\pi} \int_0^{\infty}\!ds\,{\rm Im}\Pi(s) =0 
\label{spectral1}
\quad {\rm and} \quad
\frac{1}{\pi} \int_0^{\infty}\!ds\,s \,{\rm Im}\Pi(s) =0 \ .
\label{spectral2}
\end{equation}
The approximate conformal symmetry at scales below $\Lambda^2$ will
mean, however, that the second of these integrals cannot be saturated
by a simple set of low lying   resonances. A modified second Weinberg
sum rule will emerge.

We break the integration into the region of the low lying resonances
and the region from  there up to $\Lambda^2$. (The contribution from
beyond $\Lambda^2$ will be negligible.) The scale of the lower region
is set by the dynamical mass of the fermion $\Sigma(p)$, which  has
a zero-momentum value $\Sigma(0)$, taken here to be positive, and
falls with   increasing Euclidean momentum. The dynamical mass 
is related to $F_{\pi}$ by $\displaystyle{\Sigma(0)
\approx 2\pi F_{\pi}/\sqrt{N}}$. ($\Sigma(p)$ is of course 
not a gauge invariant quantity  \cite{AMNW},
but this order of magnitude relation is true in a wide class of
gauges.) The first region extends from {\it zero} to a
continuum threshold $s_0$ which we expect to be on the order of twice
the dynamically generated fermion mass: $s_0 = O(4 \Sigma^2(0))$. In this regime, the    
integral is saturated by the
Nambu-Goldstone pseudoscalar along with massive vector and
axial-vector states. If we assume, for example, that there is only a
single, zero-width vector multiplet and a single, zero-width axial
vector multiplet, then
\begin{equation}
{\rm Im}\Pi(s)=\pi F^2_V \delta \left(s -M^2_V \right) - \pi F^2_A
\delta \left(s - M^2_A \right) - \pi F^2_{\pi} \delta \left(s \right)
\ .
\label{saturation}
\end{equation}
The zero-width approximation is valid to leading order in the large
$N$ expansion. In our case, however, taking the large $N$ limit
doesn't lead to zero width since the limit must be taken with $N_f /
N$ fixed. We  will nevertheless use this simple model for the spectrum
to  illustrate the effects of a near critical IR fixed point. 
As discussed 
above, we will take all the masses and widths
to scale to $0$ with $F_{\pi}$ at the transition.

The second region, extending from $s_0$ up to $\Lambda^2$, is
associated with the continuum, and  encodes the conformal properties
of the theory.  In this ``conformal region'' we estimate the contribution
to ${\rm Im} \Pi(s)$ by evaluating the relevant Feynman diagrams for
the vacuum polarization in the presence of a dynamically generated
fermion mass, and show that it can substantially affect the  low lying
mass spectrum through the second sum rule of Eq.~\ref{spectral2}.  We do
this by first computing $\Pi(Q^2)$ for Euclidean momentum and then
continuing   analytically.

We approximate the Euclidean computation
in this range by a single loop of fermions with dynamical mass
$\Sigma(p)$, with
Ward identities respected but with additional perturbative
corrections   neglected. The framework is similar to that of
``dynamical perturbation theory'', sometimes   employed to compute the
parameters of the low energy chiral lagrangian
\cite{HoldomTerning}. Here, however, the approximation is employed
only at the higher   momentum scales of the conformal region.
Even though this is still sub-asymptotic (below $\Lambda$), it is
plausible that the   approximation is more reliable than at low
energies where confinement sets in and the   resonances appear.

We assume that in the near-critical, broken phase of interest here,
$\Sigma(p)$ can be determined throughout the conformal region  by
solving a linearized gap equation in ladder
approximation\cite{mexico}. Instanton effects should be negligible in 
this range since large instantons (of order the inverse dynamical mass) 
are most important in chiral symmetry breaking.
The running coupling $\alpha(p)$ falls
slowly from $\alpha_*$   throughout the conformal region. The form of
$\Sigma(p)$ depends on whether $p$ is below or above the scale
$\Lambda_c$ at which $\alpha(p)$ passes through
$\alpha_c$. This scale approaches $0$ at   the transition as
$\Lambda_c /  \Lambda \sim (\alpha_* - \alpha_c )^{1/b\alpha_*}$,
where b is the coefficient of the first order term in the $\beta$
function. Below $\Lambda_c$, the solution can be written in the
approximate form
\begin{equation}
\Sigma(p)  = {\Sigma(0)^2\over p } \sin \left( \int^{p}_{O(\Sigma(0))}
{dk\over k}\sqrt{\frac{\alpha(k)}{\alpha_c}-1}+ \phi\right) \ .
\label{sin}
\end{equation}
We have taken the lower limit of integration to be of order
$\Sigma(0)$ where nonlinearities enter and change the form of the
solution. We have dropped terms explicitly involving derivatives of
$\alpha(k)$ since the coupling is near the fixed point in this regime.
Note that the sin function remains positive providing the argument 
is less than $\pi$.  The
non vanishing positive phase $\phi$ ($ < \pi$) insures that 
for momenta of order 
$\Sigma(0)$ the dynamical mass is non zero and of order $\Sigma(0)$.

The character of the solution changes above $\Lambda_c$. It is a
positive definite monotonically decreasing function, continuously
connected to the lower solution (Eq.~\ref{sin}) at $\Lambda_c$. We
have not derived a general closed form in this range, but a
qualitatively correct form can be obtained by   again
neglecting terms explicitly involving derivatives of $\alpha(p)$.
$\Sigma(p)$ can then be written in the form
\begin{equation}
\Sigma (p)=A\frac{\Sigma(0)^2}{p} \sinh \left( {\delta -
\int^{p}_{\Lambda_c} {dk\over k}\sqrt{1-\frac{\alpha(k)}{\alpha_c}}}
\right) \ ,
\label{sinh}
\end{equation}
where $A$ and $\delta$ are two positive 
definite constants of order unity. 
With $\delta$ large enough, $\Sigma(p)$ will be
positive even at the   upper end of this region ($p = \Lambda$). 

Imposing the continuity of $\Sigma(p)$ at $p=\Lambda_c$ and 
using the fact that $\alpha(k) \rightarrow
\alpha_*$ for small k leads to the
critical behavior $ \log (\Lambda_c / \Sigma(0)) \sim
{\left(\alpha_*/\alpha_c - 1\right)^{-1/2}}$, and therefore also to $ 
\log (\Lambda / \Sigma(0)) \sim {\left(\alpha_*/\alpha_c -
1\right)^{-1/2}}$.

We next use these results to compute $\Pi(Q^2)$
for Euclidean momentum and then derive   the form of ${\rm Im} \Pi(s)$
throughout the conformal region by analytic continuation. To do this,
we make a further simplification that does not change the
qualitative behavior and allows the integrations to be done
analytically. In the region below $\Lambda_c$, we take   $\alpha(p)$
to be constant and equal to the fixed point value $\alpha_*$ ($ >
\alpha_c$). The absorptive part ${\rm Im} \Pi(s)$ then takes the
form
\begin{equation}
{\rm Im} \Pi (s) = N  \frac{9\Sigma(0)^4}{16\,s^2\pi^2} \sinh
\left(\eta_{*}\pi\right) \sin \left( \eta_{*}\, {\rm ln} \frac{s}{s_0}
+ 2\phi \right)\
\label{im1}
\end{equation} 
for $\displaystyle{s_{0} <s<\Lambda_c^2}$, 
where
$\eta_{*}=\left({\alpha_*}/{\alpha_c}-1\right)^{1/2}$.

In the region well above $\Lambda_c$, we take $\alpha(p)$ to be
constant and equal to $\alpha_{\Lambda} \equiv \alpha(\Lambda) \approx
0.78\alpha_*$ ($ < \alpha_c$). The argument of the sinh 
(Eq.~{\ref{sinh}}) then becomes $\delta -  
\eta_{\Lambda}{\rm ln} (p/{\tilde{\Lambda}})$, 
where $\Lambda_c < \tilde{\Lambda}< p < \Lambda$ and 
$\eta_{\Lambda}=\left(1 - {\alpha_{\Lambda}}/{\alpha_c}\right)^{1/2} 
\simeq 0.47$. The ratio $\tilde{\Lambda}/\Lambda$ is non zero 
in the limit $\alpha_{\star} \rightarrow 0$. The absorptive part
${\rm Im}\Pi(s)$ is then 
\begin{equation}
{\rm Im} \Pi (s)= -N A^2\frac{\Sigma(0)^4}{4\, s^2\pi^2} \sin
\left(\eta_{\Lambda}\pi\right)\left[ e^{2\delta- \eta_{\Lambda}\,{\rm
ln}\frac{s}{\tilde{\Lambda}^2}} \gamma \left(\eta_{\Lambda}\right) -
e^{-2\delta + \eta_{\Lambda}\,{\rm ln}\frac{s}{\tilde{\Lambda}^2}} \gamma
\left(-\eta_{\Lambda}\right)\right] \ ,
\label{im2}
\end{equation}
where it can be shown that 
$\displaystyle{\gamma(-\eta_{\Lambda})/\gamma(\eta_{\Lambda})\le 1}$
for any $\displaystyle{0\le \eta_{\Lambda} < 1}$. 
{}For $s$ approaching 
$\Lambda_c^2$, ${\rm Im}(s)$ is negative and vanishingly small.
Througout the conformal region, ${\rm Im} \Pi(s)$ behaves like
$1/s^2$ times a function that oscillates from positive to negative
in the range up to $\Lambda_c^2$, and then remains negative from
$\Lambda_c^2$ up to $\Lambda^2$. The negativity of ${\rm Im} \Pi(s)$
above $\Lambda_c^2$ (Eq.~\ref{im2}) is insured by the same condition, $2
\delta > \eta_{\Lambda}{\rm ln} (s/\tilde{\Lambda}^2)$, that 
guarantees the
positivity of $\Sigma(p)$ in this region. That ${\rm Im} \Pi(s)$
(Eq.~{\ref{im1}}) can oscillate in the region below 
$\Lambda_c^2$ even though
$\Sigma(p)$ (Eq.~{\ref{sin}}) does not,  is insured by the fact
that the argument of the sin in Eq.~{\ref{im1}} involves the
log of a squared momentum.  Note that close to the phase transition
($\eta_{*} \rightarrow 0$), the imaginary part  for
$\displaystyle{s_{0} < {s} < \Lambda_c^2}$ is suppressed relative to
the imaginary part well above $\Lambda_c$ by a factor of $\eta_*$. This
is true even at the high end of the lower   range where
$\eta_{*}\,{\rm ln}(s/s_0) =O(1)$. This suppression factor will be
compensated by the integration weight in the 
second spectral function sum rule.

Armed with this information, we now examine the spectral function sum
rules. We first note that the first sum rule (Eq.~{\ref{spectral1}})
is concentrated at low momenta. The contribution from the region from
$s_0$ to $\Lambda^2$ is suppressed relative to the overall mass scales
by the small factor $\eta_*$ or by large inverse masses. Representing
the low lying resonances as in Eq.~{\ref{saturation}} then leads to
the familiar result
\begin{equation}
F^2_V - F^2_A = F^2_{\pi}\ .
\label{1rule}
\end{equation}
A more general representation of the resonance spectrum would replace
the left hand side of this relation with a sum over vector and axial
states.

The second sum rule (Eq.~{\ref{spectral2}}) receives important
contributions from throughout the region up to $\Lambda^2$. Using
the above expressions for ${\rm Im}\Pi(s)$, we find
\begin{equation}
F^2_V M^2_V - F^2_A M^2_A =N\,9 \frac{\Sigma(0)^4}{16\,\pi^2}
\left[\cos
\left(\eta_*\, {\rm ln} \frac{\Lambda_c^2}{s_0}+ 2\phi\right)
 - \cos
2\phi  +4\, A^2  \frac{\sin
\left(\eta_{\Lambda}\pi\right)}{9\,
\eta_{\Lambda}\pi} \,
F\left({\eta_{\Lambda},\delta}\right)\right] \ ,
\label{2rule-1}
\end{equation} 
where $F\left({\eta_{\Lambda}},\delta\right)$ 
can be shown to be positive and $O(1)$.

The quantity in square brackets arises from both above and below
$\Lambda_c^2$ and its specific form depends on the approximations we
have made. For our purposes, however, it is enough to know that it is
positive and $O(1)$. This is clearly true for the third term, arising
from above $\Lambda_c^2$. The sum of the first two terms, arising from
below $\Lambda_c^2$, is also positive as long as
$2\phi  + \displaystyle{\eta_* \,
{\rm ln}(\Lambda_c/\sqrt{s_0})} > \pi$ (a wide range of  
values).
The second sum rule can then be written in the form
\begin{equation}
F^2_V M^2_V - F^2_A M^2_A = 2a\Sigma(0)^{2}F_{\pi}^2,
\label{2rule-2}
\end{equation} 
where $a$ is expected to be positive and $O(1)$, and where we have used the relation
$\displaystyle{\Sigma(0) \approx 2\pi F_{\pi}/\sqrt{N}}$. This is 
our principal result. As in  the
case of the first sum rule, a more general resonance spectrum will
lead to a left hand side with a sum over vector and axial states. In
either case, the conformal region enhances the vector piece relative to
the axial.

Combining the two sum rules, and for simplicity restricting to the
single vector and axial vector spectrum, leads to
\begin{equation}
M^2_A - M^2_V \simeq \frac{F^2_{\pi}}{F^2_A} \left[M^2_V -
2a\Sigma(0)^2 \right] \ .
\label{spectrum}
\end{equation}
The conformal region is thus expected to give a negative contribution to
the axial-vector mass difference that is not present in QCD-like
theories. Since $2a\Sigma(0)^2$ is of order $M_V^2$,
the two states may be nearly degenerate or the vector-axial mass
pattern may even be inverted with respect to QCD. Note that this
is an asymptotic result as the critical value of $N_f$ is 
approached from below. In this limit, the overall
scale $\Sigma(0)$ is of course vanishing relative to the intrinsic
scale $\Lambda$. Equation \ref{spectrum} says that the splitting
relative to this overall scale is further reduced. The above discussion does not address the  
question of how
this new spectrum emerges as $N_f$ approaches the critical value.

This changed spectrum of states could have important consequences in
technicolor theories of electroweak symmetry breaking since 
the kind of near critical theory
discussed above could provide a natural framework for a walking
technicolor theory \cite{postmodern}. The $S$ parameter \cite{PT} 
represents  a very
important test for any technicolor theory. It is related to the
absorbitive part  of the vector-vector minus axial-axial vacuum
polarization as follows \cite{PT}:
\begin{equation}
S=4\int_0^\infty \frac{ds}{s} {\rm Im}\bar{\Pi}(s)= 4\pi
\left[\frac{F^2_V}{M^2_V} - \frac{F^2_A}{M^2_A} \right] \ ,
\label{s-def}
\end{equation}
where ${\rm Im}\bar{\Pi}$ is obtained from ${\rm Im}\Pi$ by
subtracting the Goldstone boson contribution.  By using our result for
the physical spectrum provided in  Eq.~\ref{spectrum} we have
\begin{equation}
S\simeq 4 \pi F^2_{\pi} \left[\frac{1}{M^2_V} + \frac{1}{M^2_A} -
\frac{2a \Sigma(0)^2}{M^2_V\, M^2_A}\right] \ ,
\end{equation}
where, as above,  $a= O(1)$. The last term, arising
from the conformal region, through the second spectral function
sum rule, is thus expected to be negative and of the same order as   the first two
terms. While this is a crude estimate of the $S$ parameter, it seems
clear that it is much reduced relative to QCD-like theories.

Other attempts to estimate the $S$ parameter for walking technicolor
theories have been made in the past.  In Reference \cite{SH}, based on
an exotic method of analytic continuation whose accuracy seems to be
difficult to estimate \cite{TakNagoya}, it was claimed that the 
$S$ parameter might be
negative. The present approach is more directly physical, based on the
nature of the spectrum of states in such theories.

Finally, we observe that the low energy spectrum described here can be
accommodated within the framework of an effective chiral
Lagrangian. In the approximation we have employed, the spectrum
consists of a set of Nambu-Goldstone bosons along with a multiplet of
vector and axial particles. An appropriate, parity-invariant effective
Lagrangian for this set of particles is the following
\cite{Schechter}:
\begin{eqnarray}
{\cal L} &=& \frac{1}{2} {\rm Tr} \left[D_{\mu}M
D^{\mu}M^{\dagger}\right] -\frac{1}{2} {\rm Tr} \left[F_{\mu\nu}^L
F^{L\mu\nu} + F_{\mu\nu}^R F^{R\mu\nu}\right] \nonumber \\ &&+m^2_0\,
{\rm Tr}\left[A^L_{\mu}A^{L\mu} +A^R_{\mu}A^{R\mu} \right] + h\,{\rm
Tr}\left[A^L_{\mu}M A^{R\mu} M^{\dagger}\right].
\label{lagrangian}
\end{eqnarray}
The Nambu-Goldstone bosons are encoded in the $N_f\times N_f$ meson
matrix $M$ which transforms linearly under the chiral symmetry group
$SU_L(N_f)\otimes SU_R(N_f)$, while 
$D_{\mu}M=$ $\partial_{\mu}M - ig A_{\mu}^L M + igM
A_{\mu}^R$ is the chiral covariant derivative. The
symmetry is realized nonlinearly through the constraint $M M^{\dagger}
= M^{\dagger} M = F_{\pi}^2/2$. The third and fourth terms in
the Lagrangian   reduce the chiral symmetry from local to global,
giving masses to the vector and axial vector mesons.  Using the
vacuum value  $\displaystyle{<M> =  \mbox{\boldmath $1$}
F_{\pi}/\sqrt{2} }$, we find \cite{Schechter}: 
\begin{equation}
M^2_A - M^2_V=  \frac{F_{\pi}^2}{2} \left[g^2-h\right] \ .
\label{lagdiff}
\end{equation}
Comparison with Eq.~\ref{spectrum} shows that  $h$ parameterizes 
the contribution
from the conformal region. It is important to point out
that Eq.~\ref{lagrangian} is a nonlinear effective Lagrangian for
the simplified spectrum we have considered, useful only in the broken
phase.

In this letter we have shown that the ordering pattern for
vector-axial hadronic states in $SU(N)$ vector-like gauge theories
close to a conformal transition need not be the same as predicted  in
QCD-like theories. To show this, we employed spectral function
sum rules, the known asymptotic behavior determined by asymptotic
freedom, and the fact that these theories contain an extended
``conformal region'' below the asymptotic regime. 
A simple description of the conformal  
region was used to argue that it leads to a reduced and possibly even inverted 
vector-axial mass splitting. Possible consequences for technicolor
theories were discussed.

\begin{center}
{\bf Acknowledgments}
\end{center}
\noindent
We thank R.D. Mawhinney, N. Kitazawa, J. Terning and L.C.R. Wijewardhana for 
helpful discussions. 
This work has been partially supported by the US Department 
of Energy under grant DE-FG02-92ER-40704.

\end{document}